\documentclass{article}
\usepackage{tex-style/spconf}
\usepackage{cite}
\usepackage{amsmath,amssymb,amsfonts}
\usepackage{algorithmic}
\usepackage{graphicx}
\usepackage{textcomp}
\usepackage{xcolor}
\usepackage{amsmath}
\usepackage[]{algorithm2e}
\usepackage{diagbox}
\usepackage{todonotes}
\usepackage{blindtext}
\usepackage{caption}
\usepackage{hyperref}

\SetAlCapSkip{0em}
\title{Encoding time and energy model for SVT-AV1 based on video complexity}
\name{
    \parbox{\linewidth}{\centering
    Lena Eichermüller$^{\star}$ 
    \qquad Gaurang Chaudhari$^{\dagger}$
    \qquad Ioannis Katsavounidis$^{\dagger}$  
    \qquad Zhijun Lei$^{\dagger}$ \quad
    \qquad Hassene Tmar$^{\dagger}$
    \qquad Christian Herglotz$^{\star}$
    \qquad André Kaup$^{\star}$ 
    }}

\address{$^{\star}$ Multimedia Communications and Signal Processing \\ 
    Friedrich-Alexander-Universität Erlangen-Nürnberg, Erlangen, Germany \\
    $^{\dagger}$ Meta \\ California, USA}
      
\begin{document}

\maketitle
\begin{abstract}
    The share of online video traffic in global carbon dioxide emissions is growing steadily. To comply with the demand for video media, dedicated compression techniques are continuously optimized, but at the expense of increasingly higher computational demands and thus rising energy consumption at the video encoder side. In order to find the best trade-off between compression and energy consumption, modeling encoding energy for a wide range of encoding parameters is crucial.
    We propose an encoding time and energy model for SVT-AV1 based on empirical relations between 
    the encoding time and video parameters as well as encoder configurations. Furthermore, we model the influence of video content by established content descriptors such as spatial and temporal information. 
    We then use the predicted encoding time to estimate the required energy demand 
    and achieve a prediction error of 19.6\% for encoding time and 20.9\% for encoding energy.
\end{abstract}

\section{Introduction}
Research activities in video compression have witnessed remarkable growth in response to the rising demand for high-quality video content across diverse digital platforms such as Netflix and TikTok. As video-on-demand services, social media networks, and video-conferencing applications continue to shape our digital interactions, the optimization of video encoding has become important not only in terms of compression efficiency but also regarding their energy consumption~\cite{herglotz2019decoding}. In 2015, the global greenhouse gas emissions caused by video streaming were $1.3 \%$~\cite{stephens2021carbon} and are expected to increase further due to increased video traffic~\cite{cisco2020cisco}.
To comply with the growing demand for high-resolution content, advances in video codecs focus on optimizing compression performance, i.e. reducing the size of the encoded bitstream while maintaining visual quality of the reconstructed video at the end-user's device.
The emerging video coding standard AOMedia Video 1 (AV1), released in 2018 by the Alliance of Open Media~\cite{chen2018av1}, achieves higher bitrate savings compared to its predecessor VP9~\cite{mukherjee2013latest}, however, at the cost of an increased encoding complexity~\cite{grois2018performance}.
Some actions have been made in minimizing encoding complexity by SVT-AV1 that mitigates some of the complexity overhead such as maximizing CPU utilization for multicore processing~\cite{kossentini2020svt, wu2021towards, cobianchi2022enhancing}.

Before any actions towards minimizing the carbon dioxide emission of video encoding can be made, influences of encoding configuration as well as video parameters on energy consumption during encoding have to be studied first. 
Work on estimating encoding energy for AV1 or SVT-AV1 has not been done yet, only for other video encoders.
Rodr{\'\i}guez-S{\'a}nchez et al.~\cite{rodriguez2015time} proposed an encoding time prediction for intra-only coding for high-efficiency video coding (HEVC). The energy estimate $\hat{E}_{\mathrm{enc}}$ is given by
\begin{equation}
    \hat{E}_{\mathrm{enc}} = P_{\mathrm{avg}} \cdot t_{\mathrm{enc}}
\end{equation}
and depends on the average power $P_{\mathrm{avg}}$ and the processing time $t_{\mathrm{enc}}$, that was estimated using the quantization parameter (QP). Mid-range QPs lead to larger prediction error, which Ramasubbu et al.~\cite{ramasubbu2022modeling} minmized by introducing a constant offset energy $E_0$:
\begin{equation}
    \label{eq:energy-time_lit}
    \hat{E}_{\mathrm{enc}} = E_0 + P\cdot t_{\mathrm{enc}}.
\end{equation}
Here, $P$ corresponds to the mean processing power, with $t_{\mathrm{enc}}$ and $\hat{E}_{\mathrm{enc}}$ again corresponding to the encoding time and estimated energy for HEVC, respectively. 
Furthermore, Ramasubbu et al. estimated HEVC encoding energy by exploiting bit stream features~\cite{ramasubbu2022bitstream}.

An encoding time model for HTTPS Adaptive Streaming (HAS) using HEVC was proposed by Amirpour et al. using spatiotemporal features from the video complexity analyzer (VCA)~\cite{amirpour2022light}.
Until now, only transcoding time estimators for AV1 exist, i.e. Liapin et al. propose an H.264/H.256 to AV1 transcoding using object tracking~\cite{liapin2018fast}.
Those two models have one thing in common: they obtain at one point information about the content of the video scene.
Because of this observation, we also adopt content information for our proposed model.





This paper contributes the following: 
First, we verify the energy-time correlation for SVT-AV1 and observe that we can model encoding energy from processing time. 
We then propose a single-core encoding time model based on video parameter and encoder configurations such as the preset. We study the influence of the video content by testing common content descriptors from the literature. We then use the linear model to obtain the expected energy consumption to verify that we can use any time model to estimate encoding energy.

In Sec.~\ref{sec:energy_model}, we define an encoding energy model for SVT-AV1 that depends on the processing time required for encoding a video sequence. Details about tested content descriptors are explained in Sec.~\ref{sec:content}. In Sec.~\ref{sec:evaluation}, we first verify the linear relationship between processor time and energy demand for the SVT-AV1 encoder. Then, we evaluate the encoding time model without content information and with the tested content descriptors. Finally, we give the estimation errors for combining time and energy model.
\vspace{-5pt}
\section{High-Level Encoding Energy Model for SVT-AV1}\label{sec:energy_model}
Using the energy-time correlation from~\cite{ramasubbu2022modeling}, we develop an encoding time model depending on the following parameters: preset, constant rate factor (CRF), and number of intra frames. We define the complexity $\hat{E}_{\mathrm{enc,kpix}}$ as the energy demand needed for the encoding of $1000$ pixels in 8-bit by
\begin{equation}
    \hat{E}_{\mathrm{enc,kpix}} = E_0+P\cdot t_{\mathrm{enc,kpix}} \cdot \frac{W\cdot H}{1000} \cdot n_{\mathrm{frames}},
    \label{eq:energy-time}
\end{equation}
with $t_{\mathrm{enc,kpix}}$ being the encoding time per kilopixel, and $W$ and $H$ denoting the width and height of the video sequence. The number of frames in the video is denoted as $n_\mathrm{frames}$. Furthermore, $E_0$ is the idle energy and $P$ the slope. They depend on the CPU used for encoding, whereas the encoding time is influenced by the encoder configurations and the video content.
We will estimate $E_0$ and $P$ for this energy model and propose an encoding time model to estimate the required processing time.

We assume an exponential relationship between preset $p$ and encoding time, as well as a linear relationship between the number of intra frames $n_\mathrm{intra}$ and the constant rate factor $CRF$. We estimate the encoding time per kilopixel $\hat{t}_{\mathrm{enc, kpix}}$ as 
\begin{equation}
    \hat{t}_{\mathrm{enc, kpix}} =  \mathcal{C}^{\xi} \cdot n_{\mathrm{intra}}^{\delta} \cdot\frac{1}{\mathrm{CRF}} \cdot p^{\alpha} \cdot e^{\beta \cdot p + \gamma} + t_0,
    \label{eq:encoding-time-model}
\end{equation}
with $t_0$ being the offset time, $\mathcal{C}$ an initially unknown content dependency, as well as parameters to be fitted: the preset dependencies $\alpha$, $\beta$, $\gamma$, the influence of the number of intra-coded frames $\delta$, and the video content influence $\xi$.
Apart from the exponential term, we additionally introduce a polynomial dependency for the preset: $p^\alpha$. This leads to a more accurate model.
\vspace{-5pt}
\section{Content Dependency}\label{sec:content}
We assume that the content influence on expected time or energy demand is constant for one video sequence and call this the content complexity factor $\mathcal{C}$.
Increased encoding effort first comes from \textit{spatial complexity} $\mathcal{C}_{\mathrm{S}}$, that corresponds to higher structured areas in single frames as well as \textit{temporal complexity} $\mathcal{C}_{\mathrm{T}}$, which stems from moving objects. 
To quantify the content effect on the encoding time, we combine both texture and temporal complexity of the sequence and obtain the content complexity factor, which is an adaption of criticality from Fenimore et al. \cite{fenimore1998perceptual}:
\begin{equation}
    \label{eq:content_factor}
    \mathcal{C} = f_{n, s} \left( \mathcal{C}_{\mathrm{S}} \right) \cdot f_{n, t} \left( \mathcal{C}_{\mathrm{T}} \right)
\end{equation}
Here $\mathcal{C}_{\mathrm{S}}$ denotes the influence of spatial and $\mathcal{C}_{\mathrm{T}}$ the influence of temporal changes, accompanied by normalizing functions $f_{n, s}(\cdot)$ and $f_{n, s}(\cdot)$ for both spatial and temporal content complexity, accordingly.
Both temporal and spatial complexity may not be equally important for encoding complexity. By introducing normalizing function, we can study the effect on prediction accuracy by diminishing either of them.

\subsection{Spatial Complexity}
Encoding time increases if the spatial structure is highly varying compared to flat areas. In the following, various methods to quantify this spatial complexity are introduced.
Spatial information (SI) is defined as
\begin{equation}
    \mathcal{C}_{s; \mathrm{SI}}  = \mathrm{rms}_{\mathrm{space}}\left(\mathrm{Sobel}(X(t_n))\right),
    \label{eq:si}
\end{equation}
where $X(t_n)$ is the frame at index $t_n$, $\mathrm{Sobel}(\cdot)$ denotes the Sobel filter, and $\mathrm{rms}_{\mathrm{space}}\left(\cdot\right)$ corresponds to the root mean square value over all frames~\cite{itu1999subjective}. Spatial complexity from spatial information is then given by $\mathcal{C}_{s; \mathrm{SI}}$.
Furthermore, we use spatial complexity calculated by the video complexity analyzer (VCA) from Vignesh et al.~\cite{vca}. Here, 
the spatial complexity $\mathcal{C}_{s; \mathrm{VCA}}$ can be obtained by the discrete cosine transform $\mathrm{DCT}(\cdot)$, i.e.,
\begin{equation}
    \label{eq:vca_E}
    \mathcal{C}_{s; \mathrm{VCA}} = \sum_{k=0}^{C-1} \frac{H_{p,k}}{C \cdot w^2},
\end{equation}
with the block-wise texture for each frame given as
\begin{equation}
    \label{eq:vca_blocktexture}
    H_{p,k} = \sum_{i=0}^{w-1} \sum_{j=0}^{w-1} e^{| \left(\frac{ij}{w^2}\right)^2 - 1 |} | \mathrm{DCT}(i,j) |.
\end{equation}
The number of blocks per frame is denoted as $C$, $k$ is the block-address in frame $p$, with block sizes of $w \times w$. Pixel addresses inside a block are indicated by $i$ and $j$.
Spatial complexity from variance $\mathcal{C}_{s; \mathrm{var}}$ can also be estimated by the variance, where we will use the average block-based variance with 64x64 dimensional blocks:
\begin{equation}
    \label{eq:block_variance}
    \mathcal{C}_{s; \mathrm{var}} =  \sum_{k=0}^{C-1} \frac{H_{p,k}^{\mathrm{var}}}{C \cdot 64^2}.
\end{equation}
$H_{p,k}^{\mathrm{var}}$ is the variance from block $k$ in frame $p$, and $C$ the number of blocks per frame. 

\subsection{Temporal Complexity}
Movements from objects in the video scene and from the camera lead to temporal complexity of a video sequence. The metrics used in order to quantify the amount of temporal changes are introduced in the following.

First, we use $\mathcal{C}_{t; \mathrm{TI}}$, the complexity from temporal information (TI):\vspace{-5pt}
\begin{equation}
    \mathcal{C}_{t; \mathrm{TI}} = \mathrm{rms}_{\mathrm{space}}\left(X(t_n) - X(t_{n-1})\right).
    \label{eq:ti}
\end{equation}
Again, $\mathrm{rms}_{\mathrm{space}}\left(\cdot\right)$ corresponds to the root mean square value over all frames and $X(t_n)$ is the frame at index $t_n$~\cite{itu1999subjective}.
Temporal complexity from VCA $\mathcal{C}_{t; \mathrm{VCA}}$ is tested, as well, that is\vspace{-2pt}
\begin{equation}
    \label{eq:vca_h}
    \mathcal{C}_{t; \mathrm{VCA}} = \sum_{k=0}^{C -1} \frac{\mathrm{SAD} (H_{p,k} - H_{p-1,k})}{C \cdot w^2},
\end{equation}
with $\mathrm{SAD}(\cdot)$ being the sum of absolute differences and $H_{p,k}$ the block-wise texture from~\eqref{eq:vca_blocktexture}. 
The number of frames per block is $C$ and $k$ is the block-address in frame $p$ with block sizes $w \times w$~\cite{vca}.
Last, we use optical flow $\mathcal{C}_{t; \mathrm{optical~flow}}$, where the mean spatial displacement over the sequence is used, that is\vspace{-5pt}
\begin{equation}
    \label{eq:optical_flow}
    \mathcal{C}_{t; \mathrm{optical~flow}} = \frac{1}{n_{\mathrm{frames}}} \sum_t^{n_{\mathrm{frames}}} \sqrt{(u_t+v_t)^2}.
\end{equation}
The number of frames is $n_{\mathrm{frames}}$. $u_t$ and $v_t$ correspond to the horizontal and vertical displacement at time $t$, respectively, that is calculated by the optical flow equation, which is solved using the dense flow algorithm from Farnebäck et al.~\cite{farneback2003two}.
\subsection{Ultrafast Encoding}
Another complexity indicator is the time of ultrafast encoding~\cite{ramasubbu2022modeling}, i.e., encoding with preset 13 for SVT-AV1. 
By encoding with the fastest preset, the ratio of intra- and inter-coded frames is the same as for slower presets. We assume slower presets leading to a scaling of inter- (or temporal) as well as inter-coded (or spatial) frame complexity and thus a linear scaling of encoding time.
To comply with our model, we convert this time to seconds per kilopixel:\vspace{-5pt}
\begin{equation}
    \label{eq:ultrafast}
    \mathcal{C}_{s,t; \mathrm{ultrafast}} = t_{\mathrm{preset 13}} \cdot \frac{1000}{W \cdot H \cdot n_{\mathrm{frames}}}.
\end{equation}
$\mathcal{C}_{s,t; \mathrm{ultrafast}}$ denotes the ultrafast complexity and $t_{\mathrm{preset 13}}$ the time for encoding with the fastest preset. 
\vspace{-5pt}
\section{Evaluation}\label{sec:evaluation}
We evaluate the proposed model by estimating and measuring the energy consumption and processing time when encoding with SVT-AV1 on an AMD 7452 processor @2.35 GHz uing single-core processing. We study the content of $18$ sequences from the AOM Common Test Conditions~\cite{zhao2021aom} -- all sequences from classes A3 and A4 and the following six sequences from class A2: \textit{MountainBike}, \textit{OldTownCross}, \textit{PedestrianArea}, \textit{RushFieldCuts}, \textit{WalkingInStreet}, and \textit{Riverbed}.
We encode with CRF $\in \left\{ 32, 43, 55, 63\right\}$, presets $\in \left\{ 1,2, ... , 13 \right\}$, random access configuration, and the default GOP size of $\sim 5 s$.
The mean average precision error metric (MAPE) is used for comparing estimated $\hat{y}$ and measured values $y$:\vspace{-5pt}
\begin{equation}
    \label{eq:mape}
    \mathrm{MAPE}(y, \hat{y}) = \frac{1}{n} \cdot \sum_{i=1}^n \frac{\vert y_i - \hat{y}_i \vert}{y_i}.
\end{equation}
The total number of encodings is $n$ with $i$ denoting their index.
To comply with the small number of sequences, 3-fold cross-validation is used for the time model. Each fold consists of 2 sequences from each class and two folds are used for fitting and one for validation. The reported errors are the mean for each fold.

\subsection{Energy Measurement Setup}
The energy model~\eqref{eq:energy-time} is evaluated by measuring the power consumption with an external power meter.
In order to eliminate the energy overhead from background processes, subtracting the idle energy is necessary to obtain the actual energy demand as described in~\cite{herglotz2016modeling}. Therefore, we first measure the power consumption $P_{\mathrm{total}}(t)$ during encoding over a time period $T$ and subtract the integrated idle power $ \int P_{\mathrm{idle}}(t) \mathrm{d}t$ within the same interval and get the encoding energy $E_{\mathrm{enc}}$: 
\begin{equation}
    E_{\mathrm{enc}} = \int_{t = 0}^{T} P_{\mathrm{total}}(t) \quad \mathrm{d}t - \int_{t=0}^{T} P_{\mathrm{idle}}(t)\quad \mathrm{d}t.
    \label{eq:energy_measurement}
\end{equation}
Here, $t$ corresponds to the time.
In order to remove noise, we repeat each energy measurement $m$ times until the following condition is satisfied:
\begin{equation}
    2 \cdot \frac{\sigma}{\sqrt{m}} \cdot t_{\alpha_m} \left( m - 1 \right) < \beta_m \cdot E_{\mathrm{enc}}.
    \label{eq:confidence}
\end{equation}
The maximum deviation from the actual energy is then given by $\beta$, with a probability of $\alpha$. Furthermore, $t_\alpha$ is the critical value of Student's t-distribution and $\sigma$ the standard deviation.
We chose $\alpha_m = 0.99$ and $\beta_m = 0.02$, according to~\cite{herglotz2016modeling}. 
\begin{table}[t]
\captionsetup{font=small}
    \centering
    \begin{tabular}{c|c|c|c|c}
                        &  SI               & VCA E             & Variance                          & Ultrafast     \\\hline
         TI             &  25.64   & 23.23   & 25.90                   &     -         \\\hline
         VCA h          &  20.76   & \textbf{19.64 }   & 20.78          &     -         \\\hline
         Optical flow   &  25.29   & 23.84    & 25.71                    &     -         \\\hline
         Ultrafast      & -                 &   -               &   -                               & 25.59 
    \end{tabular}
    \caption{Errors in $\%$ between estimated and measured encoding time. Best complexity estimators for both spatial and temporal complexity is from VCA which is denoted in bold numbers.}
    \label{tab:content_descriptors_errors_norm}
\end{table}
\begin{table}[t]
\captionsetup{font=small}
    \centering
    \begin{tabular}{c|c|c|c|c}
                        &  SI               & VCA E       &    Variance  &   Ultrafast     \\\hline
         TI             &  26.77  & 24.48  &    26.94  &     -         \\\hline
         VCA h          &  22.16  & \textbf{20.94 }     &    22.14  &     -         \\\hline
         Optical flow   &  26.44   & 28.94  &    26.81 &     -         \\\hline
         Ultrafast      & -                 &   -         &    -               & 26.24 
    \end{tabular}
    \caption{Errors in $\%$ between estimated and measured encoding energy. Again, the best complexity estimators for both spatial and temporal complexity is from VCA (indicated in bold).}
    \label{tab:content_descriptors_errors_energy}
\end{table}
\subsection{Evaluation of Energy Model Using Measured Encoding Time for SVT-AV1}
Encoding time measurements are obtained during energy measurements by using the CPU utilization time.
The measured data is fitted on the model from~\eqref{eq:energy-time} using least-squares fitting on all data points. 
The fitted parameter values are $E_0 = 1.68 \cdot 10^{-19}$ and $P = 1.77 \cdot 10^{-2}$ with a MAPE of $2.93 \%$. 
Therefore, we can conclude that instead of estimating encoding energy, it is sufficient to estimate encoding CPU time instead.
\subsection{Encoding Time Model Using Content Information}
\begin{figure}[t]
    \centering
    \includegraphics[width=0.45\textwidth]{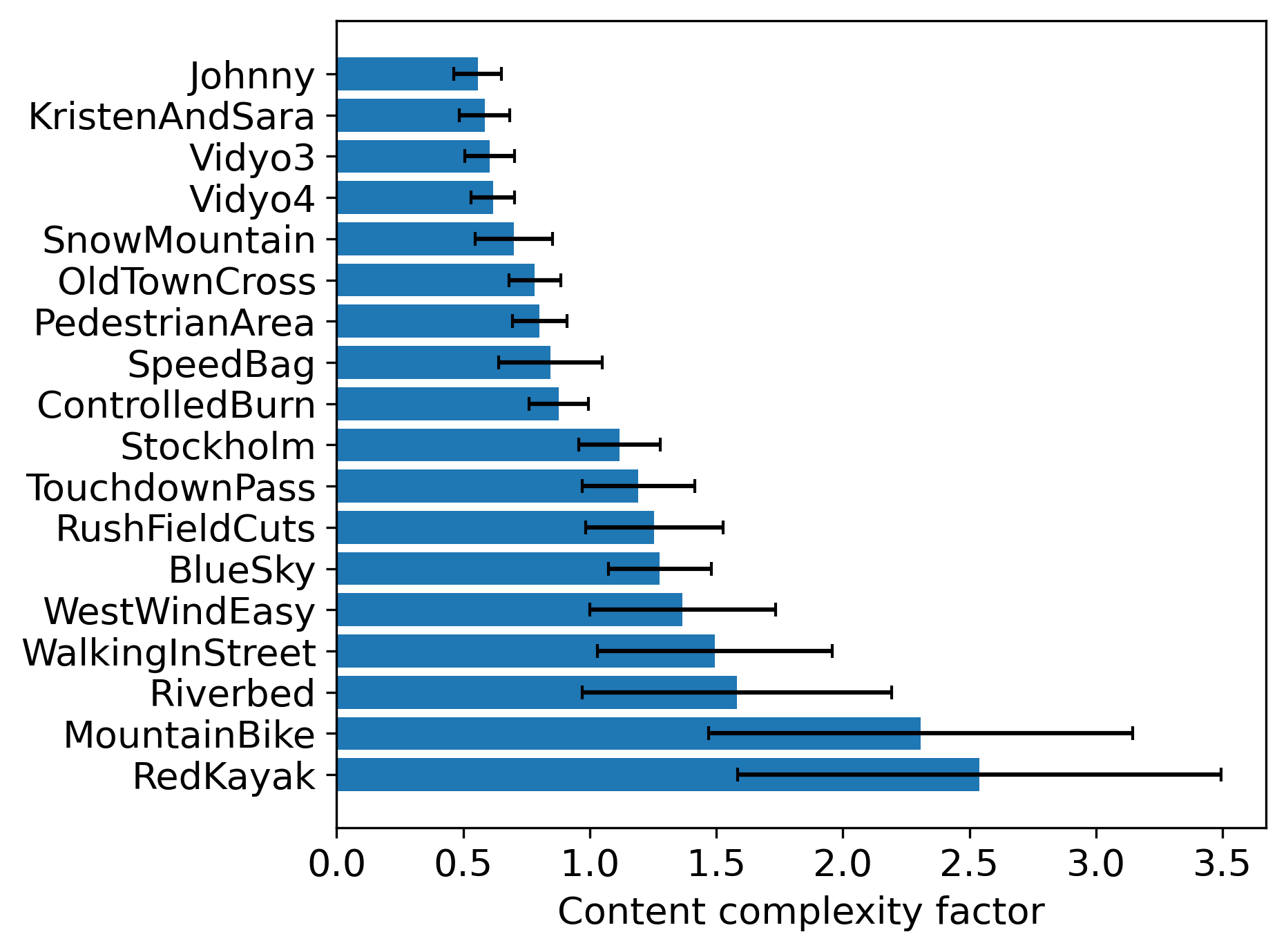}
    \caption{Optimal content factors for evaluated sequences and their standard deviations indicated by black error bars. The higher the value, the more time is required for encoding the associated sequence. 
    Larger deviations in content complexity factors are observed for high-complexity sequences.}
    \label{fig:content_factors}
\end{figure}

First, we study the influence of the video content by comparing the errors for time prediction without content information and time prediction with an oracle test by assuming that the content factor $\mathcal{C}$ is known for each sequence.
The estimated encoding time without any prior knowledge is obtained by setting the content complexity measure $\mathcal{C} = 1$ for each sequence.
This results in an estimation error of $38.44 \%$. 
For the optimal content complexity w.r.t.~\eqref{eq:encoding-time-model}, we use the fraction between real and estimated time for each sequence, averaging over all CRF--preset combinations. These are shown in Fig.~\ref{fig:content_factors}. 
The higher the content complexity factor, the higher the time or energy required for encoding, i.e. encoding the sequence \textit{RedKayak} is more complex than \textit{ControlledBurn} in terms of time and also in terms of energy. Furthermore, we observe larger deviations for higher-complexity sequences, hence describing the content influence as a constant factor for those will lead to larger prediction errors.
Assuming these theoretical optimal values for content complexity to be known and evaluating the proposed time model results in an estimation error of $17.44 \%$
, which is the lower bound when approximating constant content influence on encoding time. 

For evaluation, each spatial complexity factor $\mathcal{C}_s$ is combined with each temporal factor $\mathcal{C}_t$, as described in ~\eqref{eq:content_factor}. Normalizing functions are used for spatial complexity values, i.e., $f_{n, s} = \ln(\cdot)$ for $\mathcal{C}_{s;\mathrm{SI}}$, $\mathcal{C}_{s;\mathrm{VCA}}$, and $\mathcal{C}_{s;\mathrm{var}}$. Spatial complexity estimates have shown a lower correlation to encoding time and thus applying the logarithm diminishes their influence on content complexity. The results are listed in Tab.~\ref{tab:content_descriptors_errors_norm}. Looking at the average deviations, VCA performs best when used as spatial as well as temporal complexity with an estimation error of $19.64 \%$ 
, which is close to the best reachable error of $17.44 \%$. 
Nevertheless, using any of the tested content descriptor without content copmlexity measure significantly decreases the estimation error of $38 \%$ by almost a factor of 2.
Using criticality from~\cite{fenimore1998perceptual} as a content estimator, which is a frame-wise combination of SI and TI, leads to a MAPE of $25.64 \%$. 
VCA outperforms criticality here.
Looking at the average errors over single presets and CRFs, the highest MAPE occurs for preset 1 ($28.36 \%$) and the lowest for preset 5 ($13.47 \%$). For CRF, the highest error is at $CRF=63$ ($31.29\%$) and lowest at $CRF=43$ ($16.7\%$).
\vspace{-5pt}
\subsection{Energy Model Using Estimated Encoding Time}
The encoding energy is estimated from \eqref{eq:energy-time} with the time being approximated by \eqref{eq:encoding-time-model} and the errors are shown in Tab.~\ref{tab:content_descriptors_errors_energy}.
On average, the error increases by $1.3$ percentage points and again, the smallest prediction error is observed for combining spatial and temporal complexity from VCA resulting in an MAPE of $20.94 \%$. 
The prediction error without including any content descriptors is $39.23 \%$. 
\vspace{-5pt}

\section{Conclusion}
We observe that the encoding complexity of AV1 highly depends on the displayed content of the video. 
Using a parametric model, we can predict the encoding energy with a mean average precision error of $39\%$ without any prior information on the content and $21\%$ by incorporating content information. 
Furthermore, we have verified the correlation between processing time and encoding energy for SVT-AV1 with $3\%$ mean relative estimation error and thus we can model the consumed energy during the encoding process by its processing time. 
In future work, we will extend the model for multicore processing and study energy consumption for transmission and decoding to model the total energy consumption of video communication.

\newpage
\bibliography{references/energy_model}

\begin{thebibliography}{10}

\bibitem{herglotz2019decoding}
C.~Herglotz, A.~Heindel, and A.~Kaup, ``Decoding-energy-rate-distortion
  optimization for video coding,'' {\em IEEE Transactions on Circuits and
  Systems for Video Technology}, vol.~29, no.~1, pp.~171--182, 2019.

\bibitem{stephens2021carbon}
A.~Stephens, C.~Tremlett-Williams, L.~Fitzpatrick, L.~Acerini, M.~Anderson, and
  N.~Crabbendam, ``Carbon impact of video streaming,'' 2021.

\bibitem{cisco2020cisco}
U.~Cisco, ``Cisco annual internet report (2018--2023) white paper,'' {\em
  Cisco: San Jose, CA, USA}, vol.~10, no.~1, pp.~1--35, 2020.

\bibitem{chen2018av1}
Y.~Chen, D.~Murherjee, J.~Han, A.~Grange, Y.~Xu, Z.~Liu, S.~Parker, C.~Chen,
  H.~Su, U.~Joshi, C.-H. Chiang, Y.~Wang, P.~Wilkins, J.~Bankoski, L.~Trudeau,
  N.~Egge, J.-M. Valin, T.~Davies, S.~Midtskogen, A.~Norkin, and P.~de~Rivaz,
  ``An overview of core coding tools in the {AV1} video codec,'' in {\em Proc.
  Picture Coding Symposium (PCS)}, pp.~41--45, 2018.

\bibitem{mukherjee2013latest}
D.~Mukherjee, J.~Bankoski, A.~Grange, J.~Han, J.~Koleszar, P.~Wilkins, Y.~Xu,
  and R.~Bultje, ``The latest open-source video codec {VP9}-an overview and
  preliminary results,'' in {\em Proc. Picture Coding Symposium (PCS)},
  pp.~390--393, 2013.

\bibitem{grois2018performance}
D.~Grois, T.~Nguyen, and D.~Marpe, ``Performance comparison of {AV1, JEM, VP9},
  and {HEVC} encoders,'' in {\em Applications of Digital Image Processing XL},
  vol.~10396, pp.~68--79, SPIE, 2018.

\bibitem{kossentini2020svt}
F.~Kossentini, H.~Guermazi, N.~Mahdi, C.~Nouira, A.~Naghdinezhad, H.~Tmar,
  O.~Khlif, P.~Worth, and F.~B. Amara, ``The {SVT-AV1} encoder: overview,
  features and speed-quality tradeoffs,'' {\em Applications of Digital Image
  Processing XLIII}, vol.~11510, pp.~469--490, 2020.

\bibitem{wu2021towards}
P.-H. Wu, I.~Katsavounidis, Z.~Lei, D.~Ronca, H.~Tmar, O.~Abdelkafi, C.~Cheung,
  F.~B. Amara, and F.~Kossentini, ``{Towards much better {SVT-AV1}
  quality-cycles tradeoffs for {VOD} applications},'' in {\em Applications of
  Digital Image Processing XLIV} (A.~G. Tescher and T.~Ebrahimi, eds.),
  vol.~11842, p.~118420T, International Society for Optics and Photonics, SPIE,
  2021.

\bibitem{cobianchi2022enhancing}
G.~Cobianchi, G.~Meardi, S.~Poularakis, A.~Walisiewicz, O.~Abdelkafi, F.~B.
  Amara, F.~Kossentini, C.~Stejerean, and H.~Tmar, ``{Enhancing SVT-AV1 with
  LCEVC to improve quality-cycles trade-offs and enhance sustainability of VOD
  transcoding},'' in {\em Applications of Digital Image Processing XLV} (A.~G.
  Tescher and T.~Ebrahimi, eds.), vol.~12226, p.~122260S, International Society
  for Optics and Photonics, SPIE, 2022.

\bibitem{rodriguez2015time}
R.~Rodr{\'\i}guez-S{\'a}nchez, M.~T. Alonso, J.~L. Mart{\'\i}nez, R.~Mayo, and
  E.~S. Quintana-Ort{\'\i}, ``Time and energy modeling of an intra-only {HEVC}
  encoder,'' in {\em 2015 Visual Communications and Image Processing (VCIP)},
  pp.~1--4, 2015.

\bibitem{ramasubbu2022modeling}
G.~Ramasubbu, A.~Kaup, and C.~Herglotz, ``Modeling the {HEVC} encoding energy
  using the encoder processing time,'' in {\em 2022 IEEE International
  Conference on Image Processing (ICIP)}, pp.~3241--3245, 2022.

\bibitem{ramasubbu2022bitstream}
G.~Ramasubbu, A.~Kaup, and C.~Herglotz, ``A bit stream feature-based energy
  estimator for {HEVC} software encoding,'' in {\em 2022 Picture Coding
  Symposium (PCS)}, pp.~19--23, 2022.

\bibitem{amirpour2022light}
H.~Amirpour, P.~T. Rajendran, V.~V. Menon, M.~Ghanbari, and C.~Timmerer,
  ``Light-weight video encoding complexity prediction using spatio temporal
  features,'' in {\em 2022 IEEE 24th International Workshop on Multimedia
  Signal Processing (MMSP)}, pp.~1--6, 2022.

\bibitem{liapin2018fast}
I.~Liapin, ``Fast {H.264/H.265} to {AV1} stream transcoding using a moving
  object tracker,'' in {\em 2018 International Symposium on Consumer
  Technologies (ISCT)}, pp.~9--13, 2018.

\bibitem{fenimore1998perceptual}
C.~Fenimore, J.~Libert, and S.~Wolf, ``Perceptual effects of noise in digital
  video compression,'' in {\em 140th SMPTE Technical Conference and Exhibit},
  pp.~1--17, 1998.

\bibitem{itu1999subjective}
P.~ITU-T~RECOMMENDATION, ``Subjective video quality assessment methods for
  multimedia applications,'' 1999.

\bibitem{vca}
V.~V. Menon, C.~Feldmann, K.~Schoeffmann, M.~Ghanbari, and C.~Timmerer, ``Green
  video complexity analysis for efficient encoding in adaptive video
  streaming,'' in {\em Proceedings of the First International Workshop on Green
  Multimedia Systems}, GMSys '23, (New York, NY, USA), p.~16–18, Association
  for Computing Machinery, 2023.

\bibitem{farneback2003two}
G.~Farneb{\"a}ck, ``Two-frame motion estimation based on polynomial
  expansion,'' in {\em Image Analysis: 13th Scandinavian Conference, SCIA 2003
  Halmstad, Sweden, June 29--July 2, 2003 Proceedings 13}, pp.~363--370,
  Springer, 2003.

\bibitem{zhao2021aom}
X.~Zhao, Z.~R. Lei, A.~Norkin, T.~Daede, and A.~Tourapis, ``{AOM} common test
  conditions v2. 0,'' {\em Alliance for Open Media, Codec Working Group Output
  Document}, 2021.

\bibitem{herglotz2016modeling}
C.~Herglotz, D.~Springer, M.~Reichenbach, B.~Stabernack, and A.~Kaup,
  ``Modeling the energy consumption of the {HEVC} decoding process,'' {\em IEEE
  Transactions on Circuits and Systems for Video Technology}, vol.~28, no.~1,
  pp.~217--229, 2016.

\end{thebibliography}

\end{document}